\documentclass[cits,hyper]{PoS}
\rightline{\parbox{4cm}{\large\rm
DESY 11-024\\
Edinburgh 2010/32\\
LTH 908
}}

\usepackage{amsmath,amsthm,amsfonts,amssymb}
\usepackage[utf8]{inputenc}
\usepackage{subfig}

\newcommand{\Dlr}{\overset{\leftrightarrow}{D}}
\newcommand{\Dlrnu}{\overset{\hspace{-3mm}\leftrightarrow}{D^{\nu\}}}}
\newcommand{\Dl}{\overset{\leftarrow}{D}}
\newcommand{\Dr}{\overset{\rightarrow}{D}}

\title{Baryon Axial Charges and Momentum Fractions with $N_f=2+1$ Dynamical Fermions}

\ShortTitle{Axial Charge and Quark Momentum Fraction of Octect Baryons}

\author{M.~G\"ockeler$^{a}$, Ph.~H\"agler$^{a}$, R.~Horsley$^{b}$, Y.~Nakamura$^{a,c}$,
  D.~Pleiter$^{d}$, P.~E.~L.~Rakow$^{e}$, A.~Sch\"afer$^{a}$, G.~Schierholz$^{f}$,
  H.~St\"uben$^{g}$, \speaker{F.~Winter}$^{,b}$, J.~M.~Zanotti$^{b}$\\
        \llap{$^a$} Institut f\"ur Theoretische Physik,
                    Universit\"at Regensburg,
                    93040 Regensburg, Germany \\
        \llap{$^b$} School of Physics and Astronomy,
                    University of Edinburgh,
                    Edinburgh EH9 3JZ, UK \\
        \llap{$^c$} Center for Computational Sciences, 
                    University of Tsukuba,
                    Tsukuba, Ibaraki 305-8577, Japan\footnote{present
                      address} \\
        \llap{$^d$} John von Neumann Institute NIC / DESY Zeuthen,
                    15738 Zeuthen, Germany \\
        \llap{$^e$} Theoretical Physics Division,
                    Department of Mathematical Sciences,
                    University of Liverpool,
                    Liverpool L69 3BX, UK \\
        \llap{$^f$} Deutsches Elektronen-Synchrotron DESY,
                    22603 Hamburg, Germany \\
        \llap{$^g$} Konrad-Zuse-Zentrum f\"ur Informationstechnik Berlin,
                    14195 Berlin, Germany \\
        E-mail: \email{frank.winter@ed.ac.uk}}

\author{QCDSF/UKQCD Collaboration}

\abstract{We report on recent results of the
  QCDSF/UKQCD Collaboration on investigations of baryon structure
  using configurations generated with $N_f=2+1$ dynamical flavours of
  $O(a)$ improved Wilson fermions.
  With the strange quark mass as an additional dynamical degree of
  freedom in our simulations we avoid the need for a partially
  quenched approximation when investigating the properties of
  particles containing a strange quark, e.g. the hyperons. 
  In particular, we will focus on the nucleon and hyperon axial
  coupling constants and quark momentum fractions.}

\FullConference{The XXVIII International Symposium on Lattice Filed
  Theory, Lattice2010\\
  June 14-19,2010\\
  Villasimius, Sardinia Italy}

\begin{document}

\begin{figure}
  \begin{center}
    \subfloat[][Axial charge (unrenormalised) for N and $\Sigma$.\label{fig:gANSi}]{\includegraphics[width=.5 \textwidth]{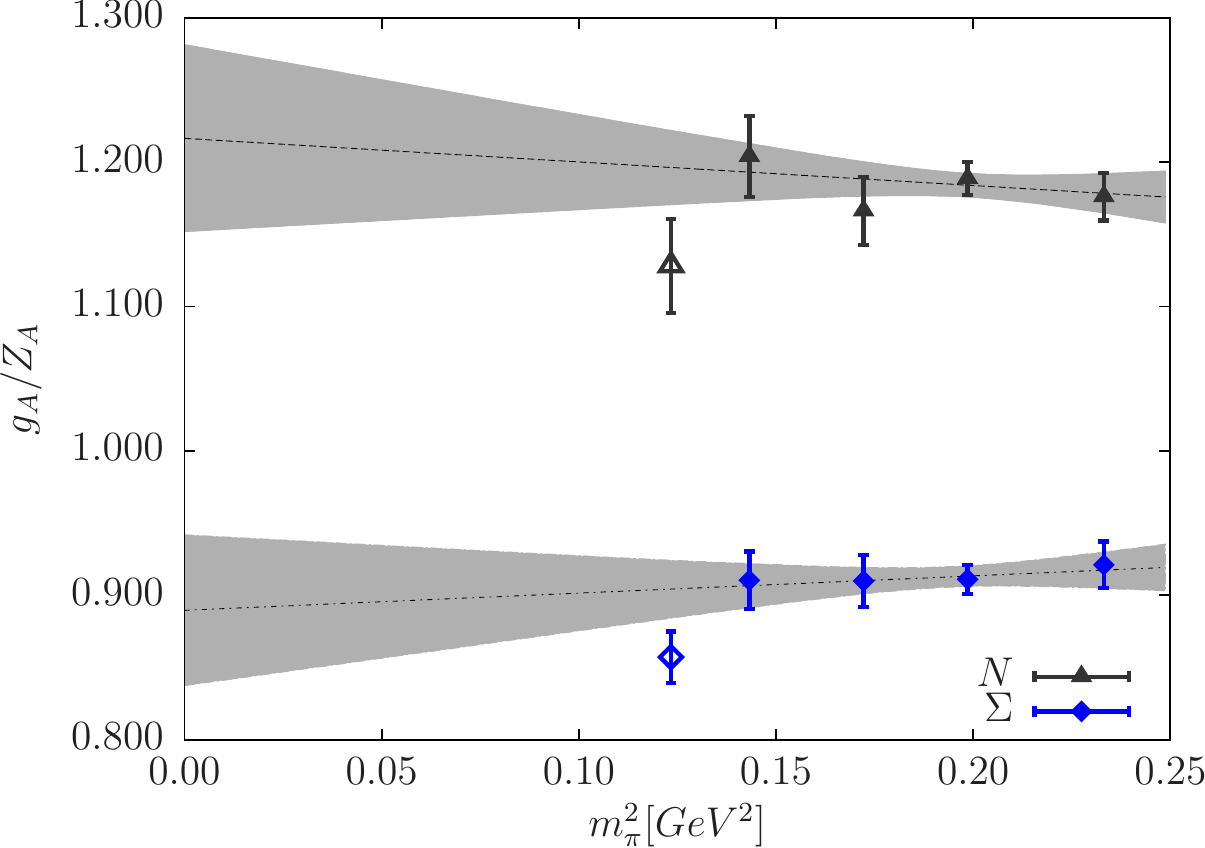}}
    \subfloat[][Axial charge (unrenormalised) for $\Xi$\label{fig:gAXi}]{\includegraphics[width=.5 \textwidth]{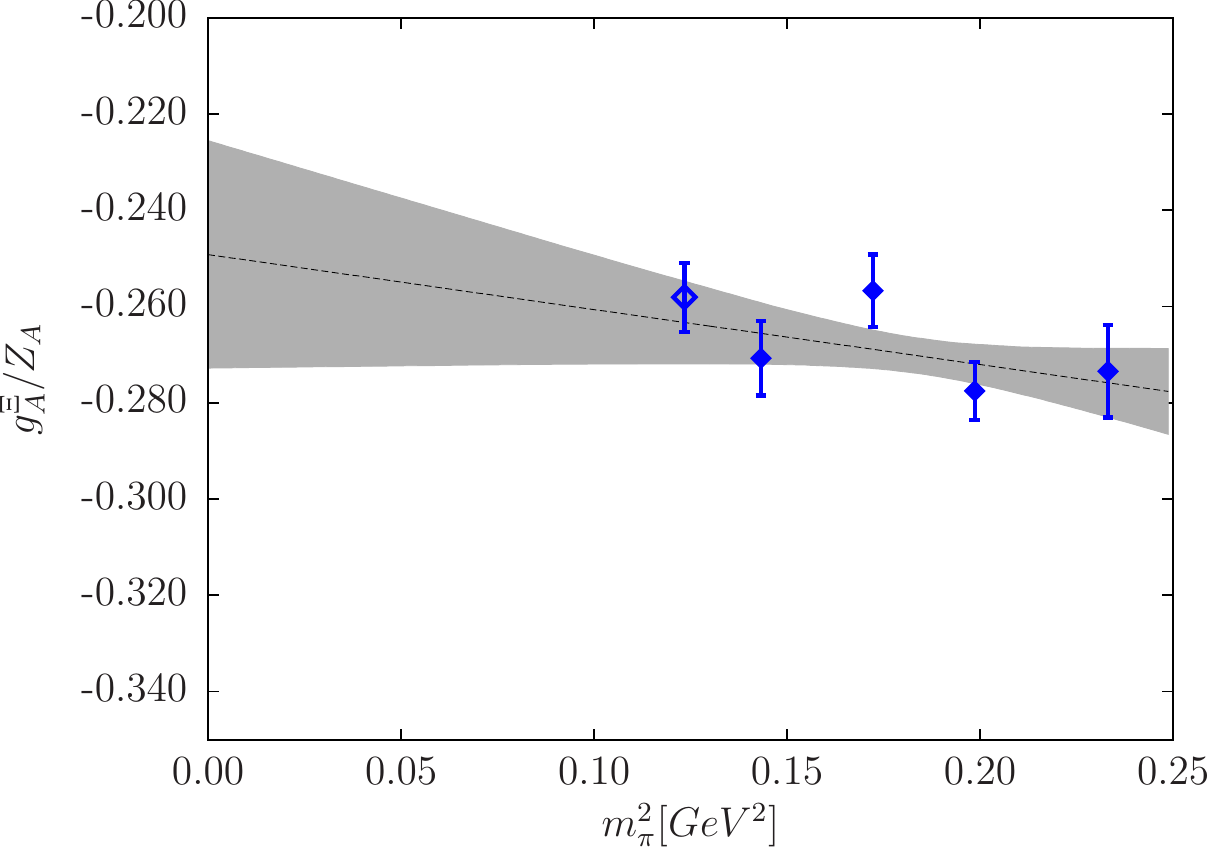}}
  \end{center}
  \caption{Unrenormalised baryon axial charge. The data points depicted with open symbols were omitted for the two-parameter linear fits.}
\end{figure}

\section{Introduction}

The nucleon axial charge governs
neutron $\beta$-decay and also provides a quantitative measure of
spontaneous chiral symmetry breaking.
It is related to the first moment of the helicity dependent quark
distribution functions, $g_A=\Delta u - \Delta d$, and has been 
studied theoretically as well as experimentally for many years.
Experiment determine its value, $g_A=1.2695(29)$, to high accuracy.
Hence it is an important quantity to study on the lattice, and since
it is relatively clean to calculate (zero momentum, isovector), it
serves as a useful yardstick for lattice simulations of nucleon
structure.

While there has been much work on the (experimentally well-known) nucleon axial charge $g_A$,
there have only been a handful of lattice investigations of the axial charge of the other
octet baryons $g_A^B$ \cite{Lin:2007ap,Sasaki:2008ha,Erkol:2009ev}, which are
relatively poorly known experimentally.
These constants are important since at leading order of SU(3) heavy
baryon chiral perturbation theory, these coupling 
constants are linear combinations of the
universal coupling constants $D$ and $F$, which enter the chiral
expansion of every baryonic quantity.

Much of our knowledge about QCD and the structure of the nucleon has
been derived from deep inelastic scattering (DIS) experiments where cross
sections are determined by its structure functions.
The DIS structure functions are related to parton distribution functions (PDF)
by asymptotic free constituents in the parton model.
Through the operator product expansion, moments of DIS structure functions
or moments of PDFs are related to matrix elements of towers of twist-2 operators (and
Wilson coefficients which are calculable in perturbative QCD).

While the quark momentum fractions of the nucleon and pion have
received much attention for many years (see, e.g.,
\cite{Hagler:2009ni} for a recent review), there have to date been no
investigations of the flavour SU(3) symmetry breaking
effects of the quark momentum fractions of the hyperons.
The obvious question that arises in this context is: {\it ``How is the
  momentum of the hyperon distributed amongst its light and strange
  quark constituents?''}

\pagebreak
We present preliminary results from the QCDSF/UKQCD Collaboration 
for the octet hyperon axial charge, $g_A^B$, and quark
momentum fraction, $\langle x\rangle_q^B$, for $B=\{N,\Sigma^+,\Xi^0\}$ 
determined with lattice QCD simulations with $N_f=2+1$ flavours of
dynamical $O(a)$ improved Wilson fermions.

\section{Simulation Details}
\label{sec:simul}

Our gauge field configurations have been generated with $N_f=2+1$
flavours of dynamical fermions, using the Symanzik improved gluon
action and nonperturbatively $O(a)$ improved Wilson fermions
\cite{Cundy:2009yy}.
We choose our quark masses by first finding the
flavour SU(3)-symmetric point where flavour singlet
quantities take on their physical values and vary the individual quark
masses while keeping the singlet quark mass
$\overline{m}_q=(m_u+m_d+m_s)/3=(2m_l+m_s)/3$ constant
\cite{Bietenholz:2010jr}.
Simulations are performed on lattice volumes of $24^3\times 48$ with
lattice spacing, $a=0.078(3)$fm.
A summary of the parameter space spanned by our dynamical
configurations can be found in Table~\ref{tab:kappas}.
\begin{table}
  \begin{center}
    \begin{tabular}{|ccccccc|}
      \hline
      $\kappa_l$ & $\kappa_s$ & $m_\pi[GeV]$ & $m_K[GeV]$ & $N\times N_T$ & $m_\pi L$ & $N_\text{meas}$ \\
      \hline
      0.120830 & 0.121040 & 0.481 & 0.420 & 24x48 & 4.63 & $~2500$ \\
      0.120900 & 0.120900 & 0.443 & 0.443 & 24x48 & 4.28 & $~4000$ \\
      0.120950 & 0.120800 & 0.414 & 0.459 & 24x48 & 3.99 & $~2500$ \\
      0.121000 & 0.120700 & 0.377 & 0.473 & 24x48 & 3.63 & $~2500$ \\
      0.121040 & 0.120620 & 0.350 & 0.485 & 24x48 & 3.37 & $~2500$ \\
      \hline
    \end{tabular}
  \end{center}
  \caption{
    Simulation parameters for $N_f=2+1$ dynamical fermions with two
    mass-degenerate light quarks and one strange quark.
    The simulation parameter $\beta$ was chosen to $\beta=5.50$ which
    corresponds to a lattice spacing of $a=0.078(3)$fm.
    \label{tab:kappas}
  }
\end{table}

\begin{figure}
  \begin{center}
    \subfloat[][Nucleon axial charge over pion decay constant\label{fig:gAfpiN}]{\includegraphics[width=.5 \textwidth]{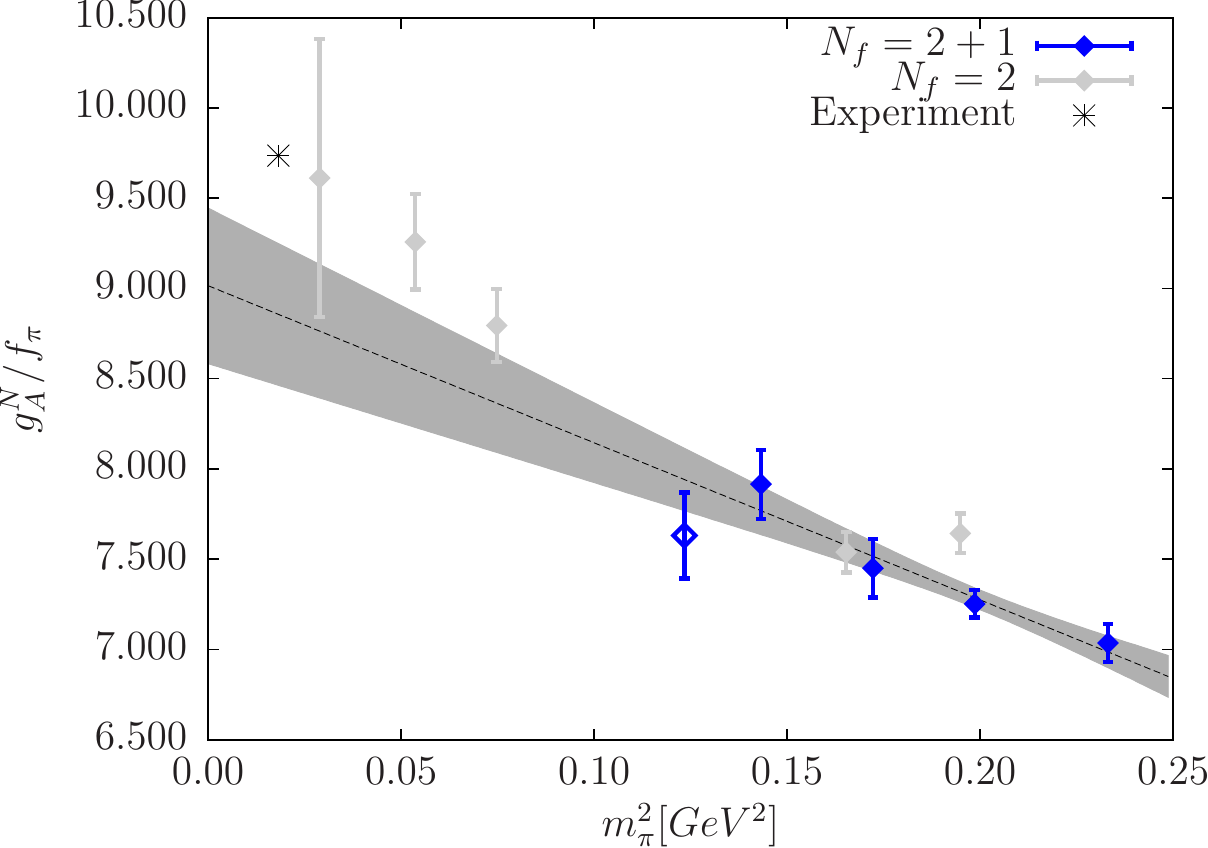}}
    \subfloat[][$g_A^\Sigma/g_A^N$\label{fig:gAr1}]{\includegraphics[width=.5 \textwidth]{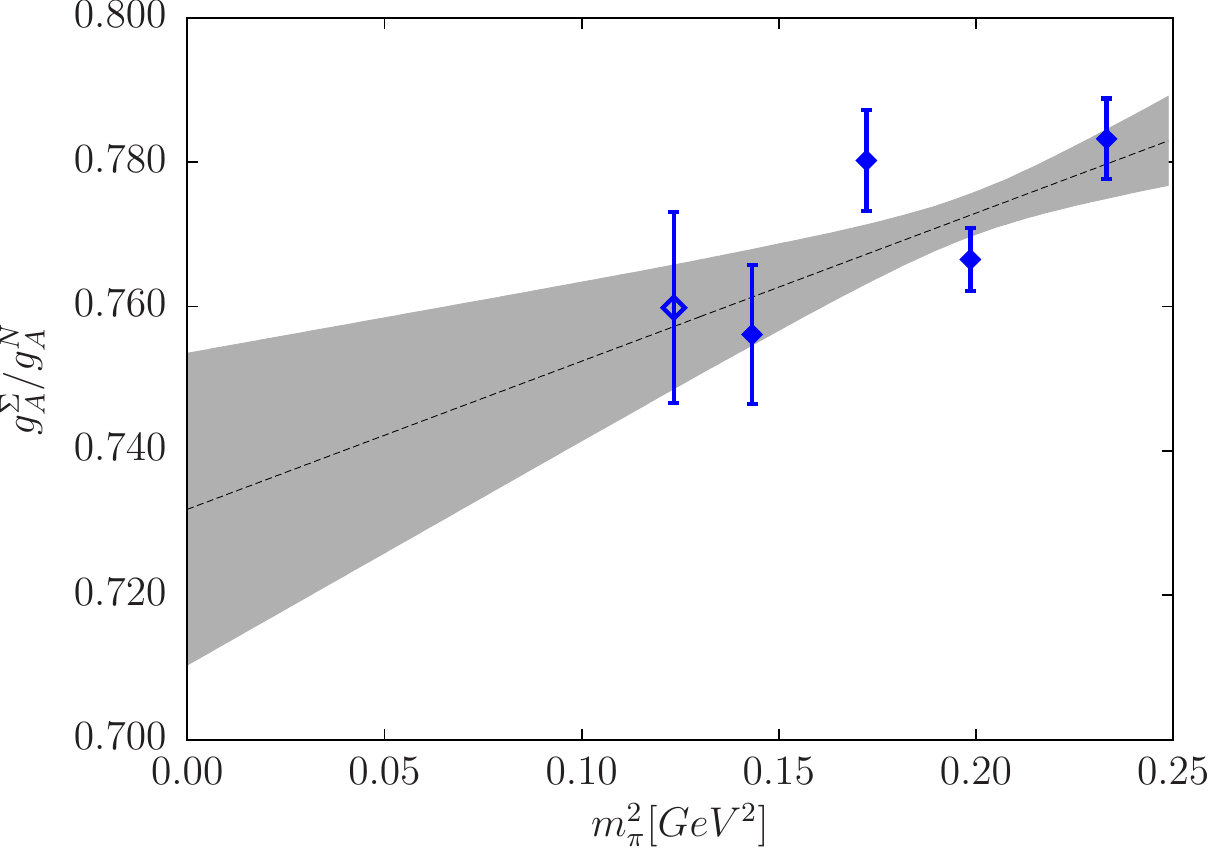}}\\
    \subfloat[][$g_A^\Xi/g_A^N$\label{fig:gAr2}]{\includegraphics[width=.5 \textwidth]{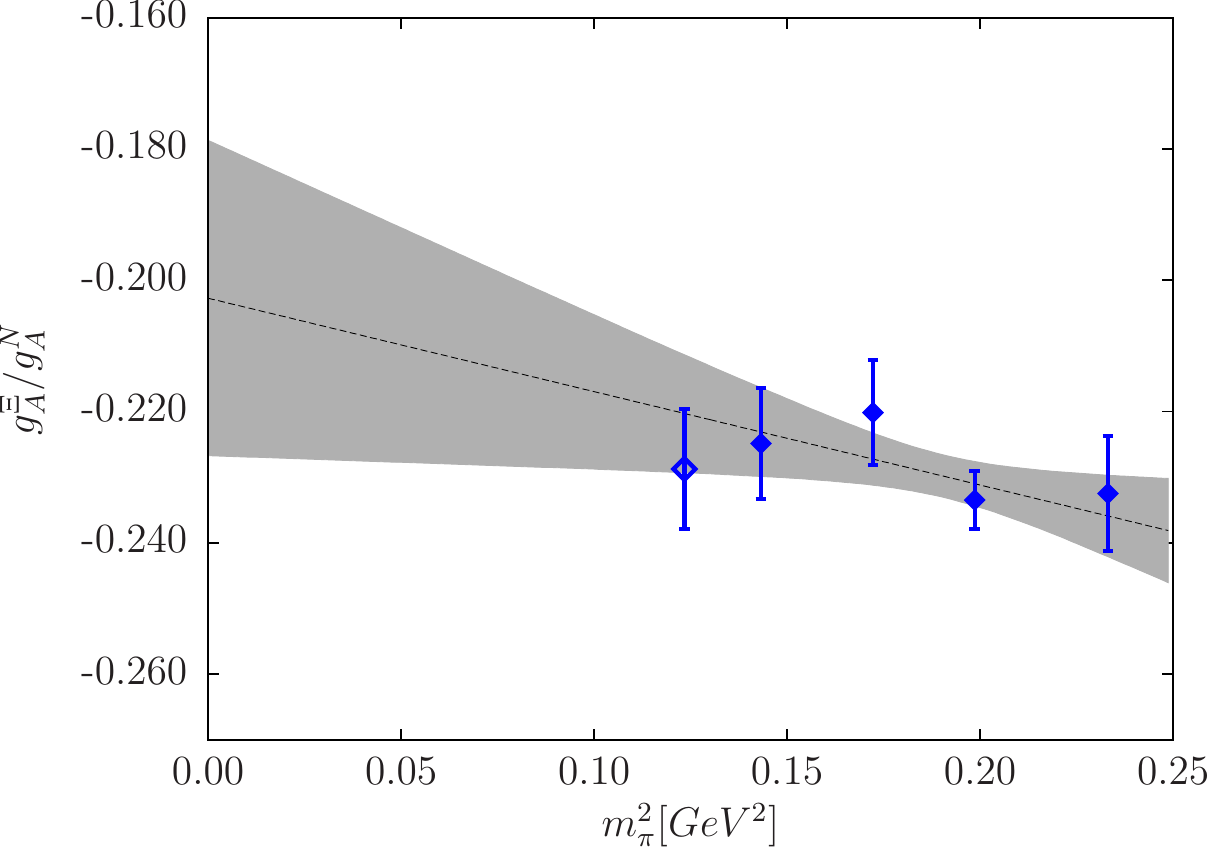}}
    \subfloat[][$(g_A^N-g_A^\Xi)/g_A^\Sigma$\label{fig:gAr3}]{\includegraphics[width=.5 \textwidth]{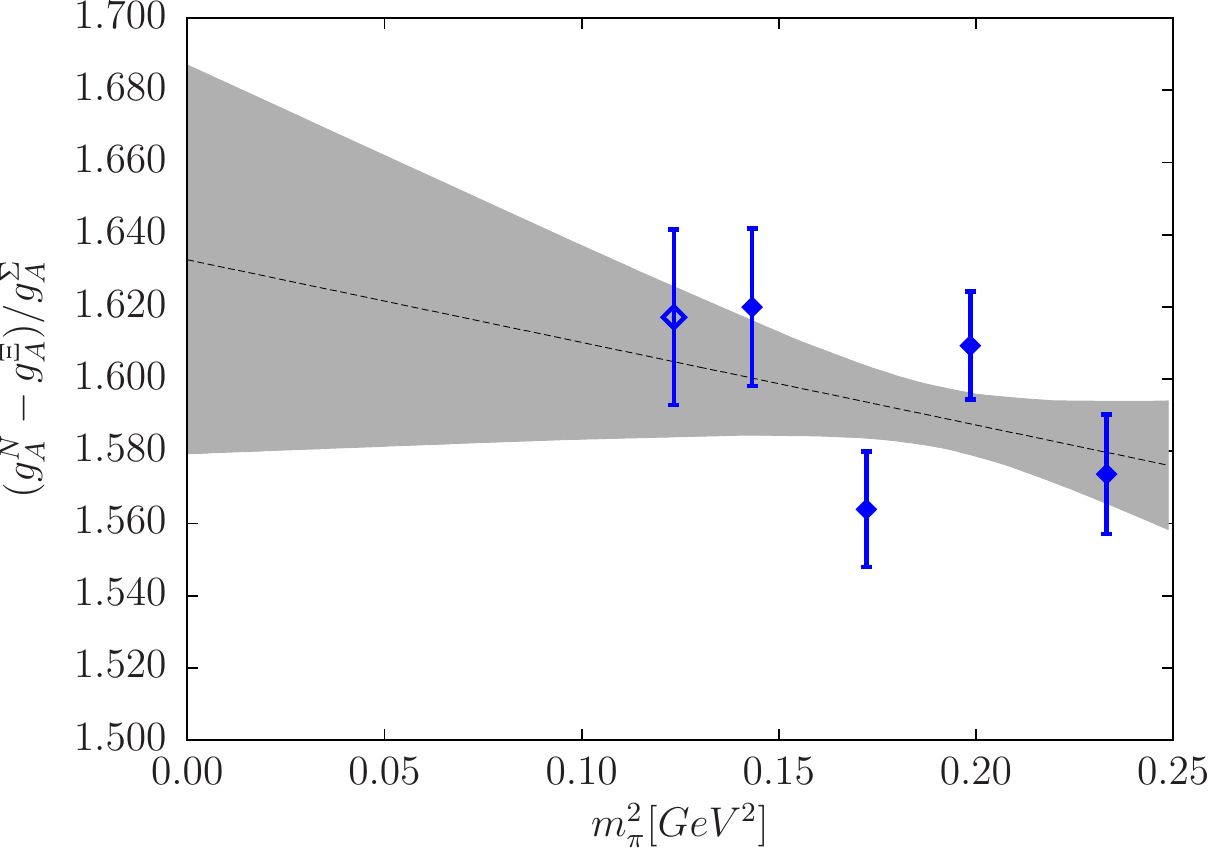}}
  \end{center}
  \caption{Ratios of unrenormalised baryon axial charge, the renormalisation constant cancels.}
\end{figure}

\section{Baryon Axial Charge $g_A^B$}

The axial charge is defined as the axial vector form factor at zero
four-momentum transfer, $g_A = G_A (0)$, which is obtained from
the matrix element for the baryon, $B$
\begin{equation}
  \langle B(p',s') | A^{u-d}_\mu | B(p,s) \rangle = 
  \overline{u}_B(p',s')
  \left[
    \gamma_\mu \gamma_5 G_A(q^2) + 
    \gamma_5 \frac{q_\mu}{2m_N} G_P(q^2) 
  \right]
  u_B(p,s)\ ,
\end{equation}
where $q=p'-p$ denotes the 4-momentum transfer and $u_B(p,s)$ is the
spinor for the baryon, $B$, with momentum $p$ and spin vector $s$ and
$G_P$ is the induced pseudoscalar form factor.
The isovector axial current is defined as $A^{u-d}_\mu=\overline u
\gamma_\mu \gamma_5 u - \overline d \gamma_\mu \gamma_5 d$ where $u$
and $d$ denote the up and down quark fields, respectively. 
We work in the limit of exact isospin invariance, i.e. $u$ and $d$
quarks are assumed to be degenerate in mass.
The states are normalised according to $\langle p',s'| p,s\rangle = (2
\pi)^3 2 p^0 \delta({\bf p}-{\bf p'})\delta_{ss'}$, we take
$s^2=-m_B^2$ and $m_B$ is the baryon mass. 
Thus the axial charge is given by the forward matrix element
%
%\begin{equation}
$
  \langle B(p,s) | A^{u-d}_\mu | B(p,s) \rangle = 2 g_A^B s_\mu.
$
%\end{equation}
%
In parton model language, the forward matrix elements of the axial
current are related to the fraction of the spin of the baryon carried
by the quarks.
Denoting by $\langle 1 \rangle^B_{\Delta q}$ the contribution of the quark, $q$,
to the spin of the baryon, $B$, one finds
\begin{equation}
    \langle B(p,s) | \overline q \gamma_\mu \gamma_5 q | B(p,s)
    \rangle = 2 \langle 1 \rangle^B_{\Delta q} s_\mu.
\end{equation}
Thus for the nucleon we write $g_A^N=\langle 1 \rangle^N_{\Delta u} -
\langle 1 \rangle^N_{\Delta d}$.

Figure \ref{fig:gANSi} (\ref{fig:gAXi}) shows the unrenormalised axial
charge for the nucleon and the $\Sigma$ ($\Xi$).
It is well known that the nucleon axial charge is sensitive to finite
size effects (FSE) \cite{Khan:2006de}. We suppose that the drop of $g_A^B$
for $B=\{N,\Sigma\}$ at the lightest pion masses is due to FSE.

With the current data we can not do better than 
a first approximation with a linear two-parameter fit to find the
unrenormalised axial charge at the physical point.

The next step would be to renormalise our results, however as yet $Z_A$
is unknown for these ensembles so we instead consider ratios where the
renormalisation constant cancels.
The first ratio we take is $g_A^N/f_{\pi^\pm}$, the nucleon axial charge over the pion decay constant, 
shown in Fig.~\ref{fig:gAfpiN}.
Since the renormalisation constants cancel in the ratio, we are able
to compare our results to the experimental value \cite{Amsler:2008zzb} and
to our $N_f=2$ results \cite{PleiterPoS2010}.
Except for the lightest pion mass, which is possibly due to FSE, the
measurements show a trend towards the experimental value and agree
very well with the $N_f=2$ results.

\begin{table}
  \begin{center}
    \begin{tabular}{|cccccc|}
  \hline
  R & $a_0$ & $a_1$ & $\chi^2/$dof & quality & value   \\
  \hline
  $g_A^\Sigma/g_A^N$ & 0.732(22) & 0.21(11) & 6.043090 & 0.002374   & 0.736(22) \\
  $g_A^\Xi/g_A^N$     & -0.203(24) & -0.14(13) & 1.299522 & 0.272662  & -0.205(24) \\
  $(g_A^N-g_A^\Xi)/g_A^\Sigma$ & 1.633(54) & -0.23(28) & 6.498347 & 0.001506 & 1.629(54) \\
  $(g_A^N+g_A^\Xi)/g_A^\Sigma$ & 1.082(45) & -0.45(23) & 0.311016 & 0.577057 & 1.074(45) \\
  \hline
    \end{tabular}
  \end{center}
  \caption{
    Ratios of the baryon axial charge in the chiral limit.
    Extrapolations to the physical point were obtained via a two-parameter linear fit model 
    $R=a_0+a_1 m_\pi^2$.
    \label{tab:gAr}
  }
\end{table}

\section{Ratios of the Axial Charge}

In the case of exact flavour SU(3) symmetry, the axial charge of the
$N$, $\Sigma$, and $\Xi$ ground states are connected by the following
linear combinations of the SU(3) constants, $F$ and $D$
\cite{Dannbom1996,Gaillard1984}
\begin{align}
  \nonumber
  g_A^N  = F + D \qquad
  g_A^\Sigma = 2F \qquad
  g_A^\Xi = D - F.
\end{align}
We consider ratios of the baryon axial charge in which the renormalisation constant cancels:
\begin{align}
  \nonumber
  \frac{g_A^\Sigma}{g_A^N} = \frac{2F}{F+D} \qquad\quad
  \frac{g_A^\Xi}{g_A^N}    = \frac{F-D}{F+D} \qquad\quad
  \frac{g_A^N-g_A^\Xi}{g_A^\Sigma} = \frac{D}{F} \qquad\quad
  \frac{g_A^N+g_A^\Xi}{g_A^\Sigma} = 1
\end{align}

\begin{figure}[t]
  \begin{center}
    \subfloat[][$(g_A^N+g_A^\Xi)/g_A^\Sigma$\label{fig:gAr4}]{\includegraphics[width=.5 \textwidth]{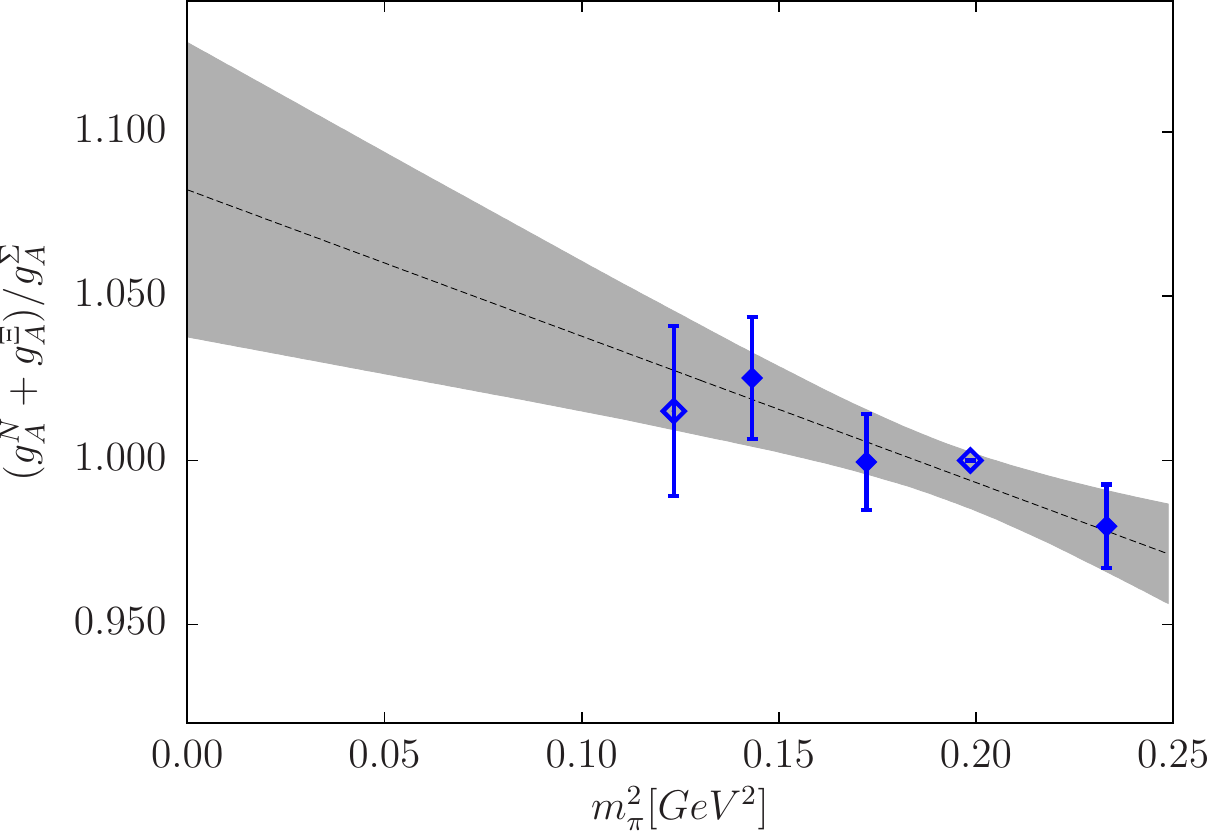}}
    \subfloat[][Unrenormalised quark momentum fractions\label{fig:xfan}]{\includegraphics[width=.5 \textwidth]{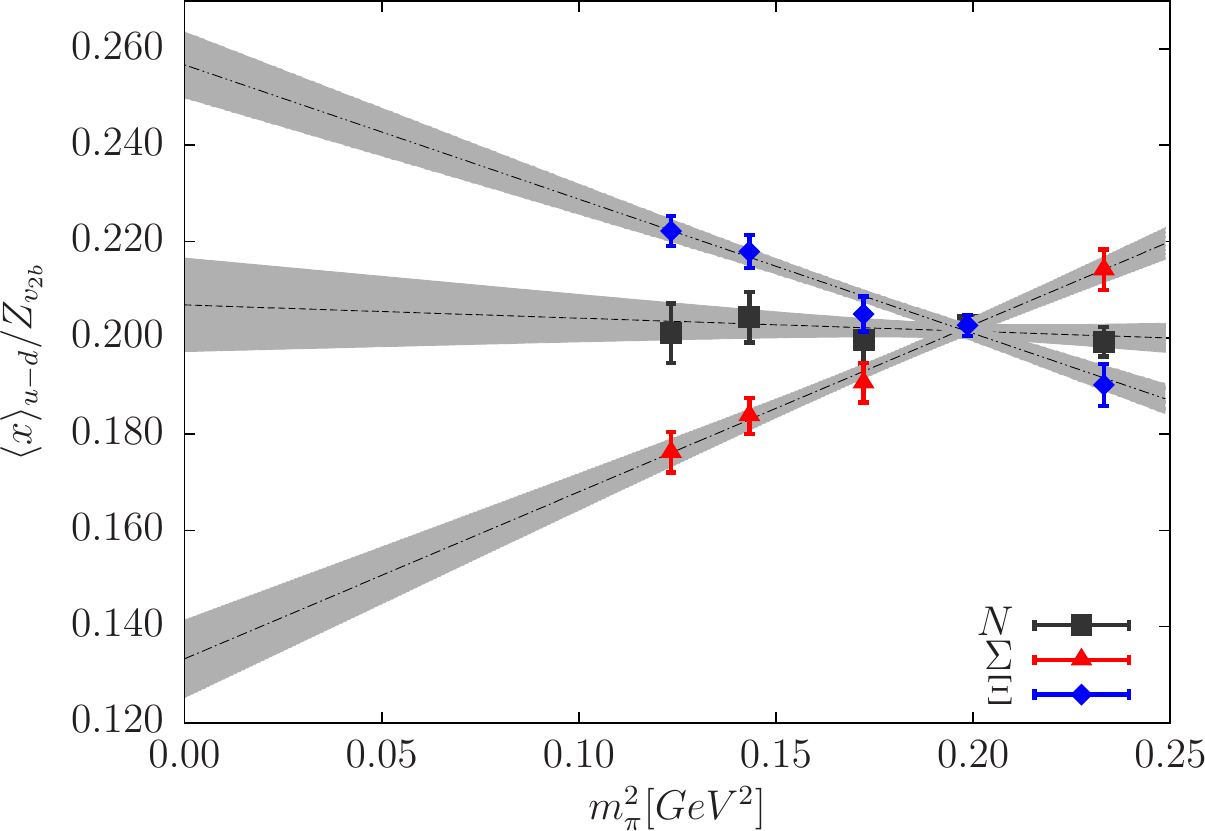}}
  \end{center}
  \caption{Left side: Ratio of baryon axial charge. Right side: Quark momentum fractions for the N, $\Sigma$, and $\Xi$.}
\end{figure}

Figs.~\ref{fig:gAr1} - \ref{fig:gAr4} show these ratios as a function of $m_\pi^2$.
We linearly extrapolate to the physical quark mass to obtain preliminary predictions 
which we collectively show in Table \ref{tab:gAr}.
These preliminary results are in good agreement with earlier
lattice  \cite{Lin:2007ap,Erkol:2009ev} and quark model
\cite{Choi:2010ty} determinations.

From a fit to the experimental data taking model independent leading 
SU(3) breaking contributions to the axial current matrix elements 
into account Savage and Walden \cite{Savage:1996zd} found the following values:
$F=0.47(7)$ and $D=0.79(10)$.
Combining the central values the following ratios are obtained:
$g_A^\Sigma/g_A^N=0.75$, $g_A^\Xi/g_A^N=-0.25$, and $(g_A^N-g_A^\Xi)/g_A^\Sigma=1.68$.
Thus, our results are in good accordance with their results.

\section{Momentum Fractions}

The first moment of a baryon's, $B$, unpolarised quark distribution
function, $q(x)$ gives the total fraction of the baryon's momentum
carried by the quark, $q$, $\langle x \rangle^B_q$.
This moment is related to the matrix element of a twist-2 operator (and a Wilson coefficient)
\begin{equation}
  \langle B(p)| \overline q \gamma^{\{\mu}i\Dlrnu q  |B(p) \rangle 
  = 2 \langle x \rangle^B_q p^{\{\mu}p^{\nu\}}\ ,
\end{equation}
where $\Dlr=(\Dr-\Dl)/2$ is the forward/backward covariant derivative.

We determine the individual connected quark contributions, $\langle x
\rangle^B_q$, and take the difference of the doubly and singly represented quark
contributions ($(D-S)$, which is $(u-d)$ in the nucleon) so that the
disconnected contributions cancel.
As in the previous section for the axial charge, the renormalisation
constant for the momentum fractions $Z_{v_{2b}}$ has not yet been
determined, and so we are not yet able to make any quantitative
predictions for the quark momentum fractions of the hyperons.

Figure \ref{fig:xfan} shows $(D-S)$ quark momentum fractions
for the nucleon $(u-d)$, $\Sigma(u-s)$, and $\Xi(s-u)$.
Here we see clear evidence for flavour SU(3)-symmetry
breaking effects as the fractions carried by the light and strange
quark fan out from the symmetric point as we decrease the pion mass.
This result indicates that the larger contribution to the baryon momentum is
carried by the (heavier) strange quark, and that this contribution
increases (and in turn, the light quark contribution decreases) as the
strange (light) quark mass is increased (decreased) towards its
physical value.

\section{Conclusions}

We have presented preliminary results from the QCDSF/UKQCD
Collaboration for the axial charge and quark momentum fraction
of the octect baryons $N, \Sigma$, and $\Xi$
from lattice QCD simulations with $N_f=2+1$ flavours of dynamical fermions.

Our results for the hyperon axial charge agree well with
earlier lattice results and show a hint of
flavour SU(3)-symmetry breaking effects.
The quark momentum fractions of the octet hyperons, on the other hand,
show strong flavour SU(3)-symmetry breaking effects, with
the heavier strange quark contributing a larger fraction to the total
baryon momentum than the light quarks.

An obvious feature that is currently lacking from these results is a
determination of the renormalisation constants for the local operators
considered here.
These calculations are now underway and will allow us to make more
quantitative predictions in the near future.

\section*{Acknowledgements}

The numerical calculations have been performed on the apeNEXT at
NIC/DESY (Zeuthen, Germany), the IBM BlueGeneL at EPCC (Edinburgh,
UK), the BlueGeneL and P at NIC (J\"ulich, Germany), the SGI ICE 8200
at HLRN (Berlin-Hannover, Germany) and the JSCC (Moscow, Russia).
We thank all institutions.
We have made use of the Chroma software suite \cite{Edwards:2004sx} and
the Bluegene codes were optimised using Bagel \cite{Boyle2005}.
This work has been supported in part by the DFG (SFB/TR 55, Hadron Physics from Lattice QCD) 
and the European Union under grants 238353 (ITN STRONGnet) and 227431 (HadronPhysics2).
JZ is supported through the UK's {\it STFC Advanced Fellowship Programme} under contract number ST/F009658/1.

\bibliography{../../bibtex/bibtex.bib}

\begin{thebibliography}{10}

\bibitem{Lin:2007ap}
H.-W. Lin and K.~Orginos,
\newblock Phys.Rev. {\bf D79}, 034507 (2009), 0712.1214.

\bibitem{Sasaki:2008ha}
S.~Sasaki and T.~Yamazaki,
\newblock Phys.Rev. {\bf D79}, 074508 (2009), 0811.1406.

\bibitem{Erkol:2009ev}
G.~Erkol, M.~Oka, and T.~T. Takahashi,
\newblock Phys.Lett. {\bf B686}, 36 (2010), 0911.2447.

\bibitem{Hagler:2009ni}
P.~Hägler,
\newblock Phys.Rept. {\bf 490}, 49 (2010), 0912.5483.

\bibitem{Cundy:2009yy}
N.~Cundy {\em et~al.},
\newblock Phys.Rev. {\bf D79}, 094507 (2009), 0901.3302.

\bibitem{Bietenholz:2010jr}
W.~Bietenholz {\em et~al.},
\newblock Phys.Lett. {\bf B690}, 436 (2010), 1003.1114.

\bibitem{Khan:2006de}
A.~Ali~Khan {\em et~al.},
\newblock Phys.Rev. {\bf D74}, 094508 (2006), hep-lat/0603028.

\bibitem{Amsler:2008zzb}
Particle Data Group, C.~Amsler {\em et~al.},
\newblock Phys. Lett. {\bf B667}, 1 (2008).

\bibitem{PleiterPoS2010}
D.~Pleiter {\em et~al.},
\newblock PoS , 153 (2010).

\bibitem{Dannbom1996}
K.~Dannbom, L.~Y. Glozman, C.~Helminen, and D.~O. Riska,
\newblock Nucl. Phys. {\bf A616}, 555 (1997), hep-ph/9610384.

\bibitem{Gaillard1984}
J.~M. Gaillard and G.~Sauvage,
\newblock Ann. Rev. Nucl. Part. Sci. {\bf 34}, 351 (1984).

\bibitem{Choi:2010ty}
K.-S. Choi, W.~Plessas, and R.~Wagenbrunn,
\newblock Phys.Rev. {\bf D82}, 014007 (2010), 1005.0337.

\bibitem{Savage:1996zd}
M.~J. Savage and J.~Walden,
\newblock Phys. Rev. {\bf D55}, 5376 (1997), hep-ph/9611210.

\bibitem{Edwards:2004sx}
SciDAC Collaboration, LHPC Collaboration, UKQCD Collaboration, R.~G. Edwards
  and B.~Joó,
\newblock Nucl.Phys.Proc.Suppl. {\bf 140}, 832 (2005), hep-lat/0409003.

\bibitem{Boyle2005}
P.~A. Boyle,
\newblock Comp.\ Phys.\ Comm. {\bf 180}, 2739 (2009).

\end{thebibliography}
\bibliographystyle{h-physrev}

\end{document}